\documentclass[times,sort&compress,3p]{elsarticle}
\journal{Journal of Multivariate Analysis}
\usepackage{amsmath,amssymb,enumerate,caption, amsfonts, dsfont, amsthm,commath}
\usepackage[english]{babel}
\usepackage[utf8]{inputenc}
\usepackage{graphicx,psfrag,epsf}
\usepackage{natbib}
\usepackage{url} 
\usepackage{booktabs, subfig, bm, paralist,mathpazo,tikz,todonotes,longtable,microtype} 
\usepackage[pdftex,colorlinks=true]{hyperref}

\newtheorem{theorem}{Theorem}
\newtheorem{assumption}{Assumption}

\newtheorem{proposition}{Proposition}

\DeclareCaptionStyle{italic}[justification=centering]{labelfont={bf},textfont={it},labelsep=colon}
\definecolor{darkblue}{rgb}{0,0,.6}
\hypersetup{citecolor=darkblue,linkcolor=darkblue,urlcolor=darkblue}
\usepackage{changes,color}

\graphicspath{{plots/}}

\newsavebox\CBox

\begin{document}

\def\spacingset#1{\renewcommand{\baselinestretch}%
{#1}\small\normalsize} \spacingset{1}

\title{\textbf{\mbox{High-dimensional functional time series forecasting: }\\An application to age-specific mortality rates}}

\author{\normalsize Yuan Gao\footnote{Postal address: Research School of Finance, Actuarial Studies, and Statistics, Australian National University, Kingsley Street, Acton ACT 2601, Australia; Telephone number: +61(2) 612 57290; E-mail: u5758483@anu.edu.au},\quad Han Lin Shang,\quad Yanrong Yang}

\address{\normalsize Research School of Finance, Actuarial Studies, and Statistics \\Australian National University}

\begin{abstract}

We address the problem of forecasting high-dimensional functional time series through a two-fold dimension reduction procedure. The difficulty of forecasting high-dimensional functional time series lies in the curse of dimensionality. In this paper, we propose a novel method to solve this problem. Dynamic functional principal component analysis is first applied to reduce each functional time series to a vector. We then use the factor model as a further dimension reduction technique so that only a small number of latent factors are preserved. Classic time series models can be used to forecast the factors and conditional forecasts of the functions can be constructed. Asymptotic properties of the approximated functions are established, including both estimation error and forecast error. The proposed method is easy to implement especially when the dimension of the functional time series is large. We show the superiority of our approach by both simulation studies and an application to Japanese age-specific mortality rates. 
\end{abstract}


\maketitle

\newpage
\spacingset{1.45}

\section{Introduction}

Functional data are considered as realizations of smooth random objects, in graphical representations of curves, images, and shapes. With the increasing capability of data storing, functional data analysis (FDA) has recieved growing attention in the last twenty years. The monographs of \cite{RS02, RS05} and \cite{RH17} provide a comprehensive account of the methodology and applications of FDA. More recent advances in this field can be found in survey papers \citep{C14, GV16,  WCM16, RGS+17, FGG17}. When infinite dimensional curves are collected sequentially, they form a functional time series (FTS) $\mathcal{X}_t(u)$, where $u \in \mathcal{I}$ and $t\in \mathbb{Z}$. 

As the popularity of functional time series grows, there has been a rapid growing body of research on functional time series modeling and forecasting. From a parametric aspect, \cite{B00} and \cite{BB07} proposed the functional autoregressive of order 1 (FAR(1)) and derived one-step-ahead forecasts that are based on a regularized form of the Yule-Walker equations. Later, FAR(1) was extended to FAR$(p)$, under which the order $p$ can be determined via a sequential hypothesis testing procedure of \cite{KR13}. \cite{KK17} proposed the functional moving average (FMA) process and introduced an innovations algorithm to obtain the best linear predictor. \cite{KKW17} extended the VAR model to the vector autoregressive moving average model. Recently, \cite{LRS17} considered long-range dependent curve time series and proposed a functional autoregressive fractionally integrated moving average model. From a nonparametric perspective, \cite{Be00} and \cite{FV06} proposed functional kernel regression to model the temporal dependence via a similarity measure defined by semi-metric, bandwidth, and kernel function. From a semi-parametric viewpoint, \cite{AV08} put forward a semi-functional partial linear model that combines both parametric and nonparametric models, and this model allows us to consider additive covariates and to use a continuous path in the past to predict future values of a stochastic process. Apart from the estimation of a conditional mean, \cite{HHR13} considered a functional autoregressive conditional heteroskedasticity model for modeling conditional variance, while \cite{AHP17} considered a functional generalized autoregressive conditional heteroskedasticity model. \cite{KRS17} considered a portmanteau test for testing autocorrelation under a functional analog of generalized autoregressive conditional heteroskedasticity model. 

To deal with infinite dimensional functions, there is a demand for efficient data reduction techniques. Functional principal component analysis (FPCA) is the most commonly used approach that serves this purpose. FPCA performs eigendecomposition on the underlying variance functions. As in multivariate principal component analysis case, most of the variance structures are captured in a vector called the principal component scores.  Some papers on FPCA include \cite{HMW06} and \cite{HH06} on theoretical properties, \cite{YMW05} for sparse longitudinal data, and \cite{LMS99} and \cite{VGS05} for some interesting applications.

The existing FPCA method has been developed for independent observations, which is a serious weakness when we are dealing with functional time series. Thus we adopt a dynamic FPCA approach \citep{PT13, HK15, RS16}, where serial dependence between the curves is taken into account. With dynamic FPCA, functional time series is reduced to a vector time series, where the individual component processes are mutually uncorrelated functional principal component (FPC) scores. 

It is often the case that we collect a vector of $N$ functions at a single time point $t$. If these $N$ functions are assumed to be correlated, multivariate functional data models should be considered. Classical multivariate FPCA concatenates the multiple functions into one to perform univariate FPCA \citep[see, e.g.,][]{RS05}. \cite{CCY14} suggested normalizing each random function as a preliminary step before concatenation. \cite{BJS11} studied a functional version of principal component analysis, where multivariate functional data are reduced to one or two functions rather than vectors. However, existing models dealing with multivariate functional data either fail to handle data with a large $N$  or are difficult to implement practically.

The main contribution of this paper is to propose a possible solution to modeling high-dimensional functional time series.  By high dimension, we allow the dimension of the functional time series $N$ to grow with the length of the observed functional time series $T$. We propose a two-fold dimension reduction technique to represent the original multivariate functional time series with a low dimension scalar time series. The proposed model has three major advantages: 
\begin{inparaenum}
\item[1)] it models $N$ functional time series simultaneously, taking the cross-covariance between the populations into account; 
\item[2)] the model avoids the problem of curse of dimensionality, which is a major problem for traditional multivariate functional models; 
\item[3)] the proposed model is conceptually simple and easy to implement.
\end{inparaenum}

Our model consists of three steps:
\begin{enumerate}
\item[1)] Dynamic FPCA is performed separately on each set of functional time series, resulting in $N$ sets of principal component scores of low dimension $p_0$ (typically less than 5);
\item[2)] The first functional principal component scores from each of $N$ sets of functional time series are combined into an $N \times 1$ vector. We fit factor models to the FPC scores to further reduce the dimension into an $r \times 1$ vector $\left(r \ll N\right)$. The same is done for the second, third, and so on until the $p_0$\textsuperscript{th} FPC scores. The vector of $N$ functional time series is reduced to $r\times p_0$ what we call factors.
\item[3)] A scalar time series model can be fitted to each factor and forecasts are produced. The forecast factors can be used to construct forecast functions.
\end{enumerate}

The proposed dimension reduction model is essentially using a matrix of small dimension $\left(r \times p_0\right)$ to represent the covariation of the original $N$ functional time series. Elements of the reduced matrix are uncorrelated and it is adequate to model each element with scalar time series models.

In the second step mentioned above, we adopt factor models that are frequently used for dimension reduction for time series data. Some early application of factor analysis to multivariate time series include \cite{A63}, \cite{PRT74} and \cite{B81}. Time series in high-dimensional settings where $N \rightarrow \infty$ together with $T$ are studied in \cite{C83}, \cite{B03}, and \cite{LYB11}. Among these, we adopt a similar approach to that considered in \cite{LYB11}, where the model is conceptually simple and the asymptotic properties are established. The reason why we use the technique is that the dimension reduction is based on the lag covariance of the data, which is suitable for time series data.

The remainder of the paper is organized as follows. In Section~\ref{RM}, more detailed background on dynamic FPCA is introduced and the two-fold dimension reduction model is proposed. In Section~\ref{AT}, asymptotic results for the proposed model are given. We present simulation studies in Section~\ref{SS}. In Section~\ref{MF}, we apply our proposed model to Japanese age- and sex-specific mortality rate data. The conclusion is presented in Section~\ref{C}, and proofs are provided in the Appendix.

\section{Methods}\label{RM}

We consider the stationary $N$-dimensional functional time series $\{\bm{\mathcal{X}}_t: t = 1,\ldots, T\}$, where $T$ is the sample size. At time $t$, $\bm{\mathcal{X}}_t = [\mathcal{X}^{1}_t(u), \ldots, \mathcal{X}^{N}_t(u)]^\top$, and each $\mathcal{X}^{(i)}_t(u)$ takes values in the space $H: = L^2(\mathcal{I})$ of real-valued square integrable functions on $\mathcal{I}$. The space $H$ is a Hilbert space, equipped with the inner product $\langle x, y\rangle: = \int_{\mathcal{I}}x(u)y(u)du$. The function norm is defined as $\|x\|:=\langle x, x \rangle^{1/2}$. We could also look at the data in another direction, where we call $\{\mathcal{X}_t^{(i)}(u): t= 1,\ldots,T$ the $i$\textsuperscript{th} population of the functional time series, and there are $N$ populations. Under our setting, both the sample size and the number of populations go to infinity, that is $N\rightarrow\infty, T \rightarrow\infty$.

The purpose is to reduce the dimension of the vector functional time series data $\bm{\mathcal{X}}_t$. Our technique consists of performing FPCA on $\mathcal{X}_t^{(i)}$ for each population in the first step, resulting in $N\times p_0$ FPC scores, and then fitting factor models in the second step, getting $r\times p_0$ factors. 

In Section~\ref{sec:2.1} and \ref{sec:2.2}, we will introduce the two models we use in our two-fold dimension reduction. In Section~\ref{sec:2.3}, estimation process combining the two steps is explained. In Section~\ref{sec:2.4}, we illustrate how functional time series forecast can be performed. 

\subsection{Dynamic functional principal component analysis}\label{sec:2.1}

For each $i \in \{1, \ldots, N\} $, we assume that $\mathcal{X}^{(i)}_t$ has a continuous mean function $\mu^{(i)}(u)$ and an auto-covariance function at lag $h$, $c^{(i)}_h(u, v)$, where
\begin{align*}
\mu^{(i)}(u)   &= \rm E[\mathcal{X}^{(i)}(u)], \nonumber\\
c^{(i)}_h(u, v)   &= \rm{cov}[\mathcal{X}^{(i)}_t(u), \mathcal{X}^{(i)}_{t+h}(v)].
\end{align*}
The long-run covariance function is defined as 
\begin{align*}
c^{(i)}(u, v) = \sum_{h = -\infty}^{\infty}c^{(i)}_{h}(u, v).
\end{align*}
Using $c^{(i)}(u,v)$ as a kernel, we define the operator $C^{(i)}$ by:
\begin{align*}
C^{(i)}(x)(u) = \int_{\mathcal{I}}c^{(i)}(u,v)x(v)dv, \quad u, v \in \mathcal{I}.
\end{align*}
For simplicity, we can also write
\begin{align}\label{eq:13}
C^{(i)} = \sum_{h= -\infty}^\infty C^{(i)}_h,
\end{align}
where $C^{(i)}_h$ is the covariance operator at lag $h$. The operator is symmetric and non-negative definite. By Mercer's theorem, the operator $C^{(i)}$ admits an eigendecomposition
\begin{align}\label{eq:14}
C^{(i)}(x) = \sum_{p =1}^{\infty} \lambda^{(i)}_p\langle x, \gamma^{(i)}_p\rangle\gamma^{(i)}_p, 
\end{align}
where $(\lambda^{(i)}_p: p \geq1)$ are the eigenvalues of $C^{(i)}$ in descending order and $\big(\gamma^{(i)}_p: p \geq 1\big)$ the corresponding normalized eigenfunctions. By Karhunen-Lo\`{e}ve theorem, $\mathcal{X}^{(i)}_t(u)$ can be represented with 
\begin{align*}
\mathcal{X}^{(i)}_{t}(u) = \sum_{p = 1}^{\infty} \beta^{(i)}_{p,t}\gamma^{(i)}_p(u),
\end{align*}
where $\beta^{(i)}_{p,t} = \int_{\mathcal{I}} \mathcal{X}^{(i)}_t(u)\gamma^{(i)}_p(u)du$ is the $p$\textsuperscript{th} FPC score at time $t$. The infinite-dimensional functions can be approximated by the first $p_0$ FPC scores:
\begin{align}\label{eq:21}
\mathcal{X}^{(i)}_{t}(u) = \sum_{p = 1}^{p_0} \beta^{(i)}_{p,t}\gamma^{(i)}_p(u) + \theta^{(i)}_{t}(u),
\end{align}
where $\theta_t^{(i)}(u) = \sum_{p=p_0+1}^{\infty}\beta_{p,t}^{(i)}\gamma_p^{(i)}(u)$ captures the remaining terms cutting from $p = p_0+1$ to $\infty$.

\subsection{Factor model}\label{sec:2.2}

With the first step dimension reduction, we now have FPC scores $\beta^{(i)}_{p,t}$, where $i = 1, \ldots, N$. Define the vector FPC score as 
\begin{align}\label{eq:4}
\bm{\beta}_{p,t} = \left(\beta^1_{p,t}, \ldots, \beta^N_{p,t}\right)^\top.
\end{align}
So  $\bm{\beta}_{p,t}$ is the vector that contains the $p$\textsuperscript{th} FPC score of all $N$ functional time series.
We consider the following factor model for each $p = 1, \ldots, p_0$. Let
\begin{align}\label{eq:5}
\bm{\beta}_{p,t} = \bm{A}_p \bm{f}_{p,t} + \bm{e}_{p,t},  \quad t = 1,\ldots T,
\end{align}
where $\bm{f}_{p,t}$ is an $r \times 1$ unobserved factor time series; $\bm{A}_p$ is an $N \times r$ unknown constant factor loading matrix. We need to fit $p_0$ factor models to the FPC scores. The factor model is similar to the model in \cite{LYB11}. The difference is that in their paper, the $\bm{e}_t$'s are assumed to be white noise with mean zero and a constant covariance matrix. In our settings, there is no model on the error term, and $\bm{e}_{p,t}$ is what is left after taking out the main explaining factors $\bm{f}_{p,t}$. To write out $\bm{e}_{p,t}$:
\begin{align*}
\bm{e}_{p,t} = \bm{A}^{\prime}_{p,t}\bm{f}^{\prime}_{p,t},
\end{align*}
where $\bm{A}'_p$ is an $N \times (N-r)$ matrix, the columns of which are orthogonal to the columns of $\bm{A}_p$; and $\bm{f}'_{p,t}$ is $(N-r)\times 1$ vector.

Combing \eqref{eq:21} and \eqref{eq:5}, the original functional time series can be modeled as 
\begin{align}\label{eq:8}
\mathcal{X}^{(i)}_{t}(u) = \sum_{p = 1}^{p_0} \left[\bm{A}_p \bm{f}_{p,t}\right]^{(i)}\gamma^{(i)}_p(u)+ \epsilon^{(i)}_{t}(u),
\end{align}
where $[\cdot]^{(i)}$ denotes the $i$\textsuperscript{th} element in the vector. Note that the $i$th element in the vector $\bm{A}_p\bm{f}_{p,t}$ is in fact $\bm{\alpha}^{(i)}\bm{f}_{p,t}$, where $\bm{\alpha}^{(i)}$ is the $i$th row in the matrix $\bm{A}_p$. The resulting dimension reduced factor $\bm{f}_{p,t}$ does not rely on $i$. The error term $\epsilon_t^{(i)}(u)$ contains the accumulated error from both steps:
\begin{align*}
 \epsilon^{(i)}_t(u) = \theta_t^{(i)}(u) + \sum_{p=1}^{p_0}[\bm{e_{p,t}}]^{(i)}\gamma_p(u).
\end{align*}

Before the estimation process is introduced, we make a few notations and definitions. Define
\begin{align*}
\bm{\Sigma}^{(p)}_{\beta} (h) = \rm{cov}(\bm{\beta}_{p,t+h}, \bm{\beta}_{p,t}), \quad
\bm{\Sigma}^{(p)}_{f} (h) = \rm{cov}(\bm{f}_{p,t+h}, \bm{f}_{p,t}), \quad
\bm{\Sigma}^{(p)}_{e}(h) = \rm{cov}(\bm{e}_{p,t+h}, \bm{e}_{p,t}),
\end{align*}
and also the cross-covariance between the factor and the error term
\begin{align*}
\bm{\Sigma}^{(p)}_{f,e}(h) = \rm{cov}(\bm{f}_{p,t+h}, \bm{e}_{p,t}), \quad
\bm{\Sigma}^{(p)}_{f,e}(-h) = \rm{cov}(\bm{e}_{p,t+h}, \bm{f}_{p,t}).
\end{align*}

Using \eqref{eq:5}, we can write the relation as:
\begin{align}\label{eq:22}
\bm{\Sigma}^{(p)}_{\beta}(h) = \bm{A}_p\bm{\Sigma}^{(p)}_{f}(h)\bm{A}_p^\top + \bm{A}_p\bm{\Sigma}^{(p)}_{f,e}(h)+ \bm{A}_p\bm{\Sigma}^{(p)}_{f,e}(-h) + \bm{\Sigma}^{(p)}_{e}(h).
\end{align}
Let 
\begin{align}\label{eq:23}
\bm{L}^{(p)} &= \sum_{h = 1}^{h_0} \bm{\Sigma}^{(p)}_{\beta}(h)\bm{\Sigma}^{(p)}_{\beta}(h)^\top,
\end{align}
where $h_0$ is a constant.

Plugging \eqref{eq:22} into \eqref{eq:23}, we have
\begin{align}\label{eq:24}
\bm{L}^{(p)} = \bm{L}^{(p)*} + \bm{E}^{(p)},
\end{align}
where 
\begin{align*}
 \bm{L}^{(p)*} = \bm{A}_p\left[\sum_{h=1}^{h_0} \{\bm{\Sigma}^{(p)}_f(h)\bm{A}_p^\top + \bm{\Sigma}^{(p)}_{f,e}(h) + \bm{\Sigma}^{(p)}_{f,e}(-h)\}\{\bm{\Sigma}^{(p)}_f(h)\bm{A}_p^\top + \bm{\Sigma}^{(p)}_{f,e}(h) + \bm{\Sigma}^{(p)}_{f,e}(-h)\}^\top\right]\bm{A}_p^\top,
\end{align*}
and 
\begin{align}\label{eq:9}
\bm{E}^{(p)} =& \bm{A}_p\left[\bm{\Sigma}^{(p)}_f(h)\bm{A}_p^\top + \bm{\Sigma}^{(p)}_{f,e}(h) + \bm{\Sigma}^{(p)}_{f,e}(-h)\right]\Sigma^{(p)\top}_e(h)\\
 &+ \bm{\Sigma}^{(p)}_e(h)\left[\bm{\Sigma}^{(p)}_f(h)\bm{A}_p^\top + \bm{\Sigma}^{(p)}_{f,e}(h) + \bm{\Sigma}^{(p)}_{f,e}(-h)\right]^\top \bm{A}_p^\top \nonumber + \bm{\Sigma}^{(p)}_e(h)\bm{\Sigma}^{(p)\top}_e(h).
\end{align}
If we perform eigendecomposition on the middle part within the square brackets of $\bm{L}^{(p)*}$, then $\bm{L}^{(p)*} = \bm{A}_p\bm{U}_p\bm{D}_p\bm{U}_p^\top\bm{A}_p^\top$, where $\bm{D}_p$ is the diagonal matrix with the first $r$ largest eigenvalues. $\bm{U}_p$ is an orthogonal matrix, so that $\bm{A}_p\bm{U}_p$ is a rotation on the matrix $\bm{A}_p$. We use $\bm{A}_p\bm{U}_p$ as the matrix $\bm{A}_p$. Thus, $\bm{L}^{(p)*} = \bm{A}_p\bm{D}_p\bm{A}_p^\top$. Let the columns of $\bm{A}_p$ be the eigenvectors of the matrix $\bm{L}^{(p)*}$ corresponding to the first $r$ largest eigenvalues in descending order. The matrix $\bm{D}_p$ is then a diagonal matrix with the first $r$ eigenvalues on its diagonal. 

\subsection{Estimation}\label{sec:2.3}

We need to estimate $\bm{A}_p$, $\bm{f}_{p,t}$ and $\gamma^{(i)}_p(u)$ in \eqref{eq:8}. In the dynamic FPCA step, the long-run covariance function $c^{(i)}(u,v)$ can be estimated by:
\begin{align*}
\widehat{c}^{(i)}(u, v) = \sum_{|h|\le q}W\left(\frac{h}{q}\right)\widehat{c}^{(i)}_{h}(u,v),
\end{align*}
and the covariance operator by: 
\begin{align*}
\widehat{C}^{(i)}(x)(u) = \int_{\mathcal{I}}\widehat{c}^{(i)}(u,v)x(v)dv, 
\end{align*}
or
\begin{align}\label{eq:7}
\widehat{C}^{(i)} = \sum_{|h|\le q} W\left(\frac{h}{q}\right)\widehat{C}^{(i)}_h,
\end{align}
where
\begin{align*}
\widehat{c}^{(i)}_{h}(u,v) = \left\{ \begin{array}{ll}\frac{1}{T-h}\sum_{j = 1}^{T-h}\left[\mathcal{X}^{(i)}_{j}(u)-\overline{\mathcal{X}}_j(u)\right]\left[\mathcal{X}_{j+h}(v)-\overline{\mathcal{X}}(v)\right], \quad h\geq 0 \nonumber\\
\frac{1}{T-h}\sum_{j = 1-h}^{T}\left[\mathcal{X}^{(i)}_{j}(u)-\overline{\mathcal{X}}_j(u)\right]\left[\mathcal{X}_{j+h}(v)-\overline{\mathcal{X}}(v)\right], \quad h < 0
\end{array}\right. .
\end{align*}
Here, $W(\cdot)$ is a weight function with $W(0) =1, W(u)= W(-u), W(u) =0$ if $|u|>m$ for some $m>0$, and $W$ is continuous on $[-m, m]$. Some possible choices include Bartlett, Parzen, Tukey-Hanning, quadratic spectral and flat-top functions \citep{A91, AM92}. In this paper, we use $W(h/q) = 1-|h|/q$, where $q$ is a bandwidth parameter. Conditions will be imposed on $q$ in Section~\ref{AT}. 

By performing eigendecomposition on $\widehat{C}^{(i)}$, we can estimate empirical eigenfunctions $\widehat{\gamma}^{(i)}_p(u)$ and  the empirical FPC scores $\widetilde{\beta}^{(i)}_{p,t}  = \int_{\mathcal{I}} \mathcal{X}^{(i)}_{t}(u)\widehat{\gamma}^{(i)}_p(u)du$, calculated by numerical integration.

The estimates $\widetilde{\beta}^{(i)}_{p,t}$ are combined into a vector
\begin{align}\label{eq:6}
\widetilde{\bm{\beta}}_{p,t} = \left(\widetilde{\beta}_{p,t}^1\ldots, \widetilde{\beta}_{p,t}^N\right),
\end{align}
 and fitted to a factor model. The estimation of latent factors for high-dimensional time series can be found in \cite{LYB11}. The idea of estimation is that in the previously defined matrix $\bm{L}^{(p)} = \bm{L}^{(p)*} + \bm{E}^{(p)}$ in~\eqref{eq:24}, when the term $\bm{E}^{(p)}$ related to error covariance is small, such that $\bm{L}^{(p)*}$ is close to $\bm{L}^{(p)}$, we can use the eigendecomposition of $\bm{L}^{(p)}$ to estimate the eigendecomposition of $\bm{L}^{(p)*}$. Details can be found in Appendix.

Then, a natural estimator for $\bm{A}_p$ can be found by performing eigendecomposition on an estimated version of $\bm{L}^{(p)}$. It is defined as $\widehat{\bm{A}}_p = (\widehat{\bm{a}}_{p,1},\ldots, \widehat{\bm{a}}_{p,r})$, where $\widehat{\bm{a}}_{p,j}$ is the $j$\textsuperscript{th} eigenvector of $\widehat{\bm{L}}^{(p)}$, and 
\begin{align}\label{eq:10}
\widehat{\bm{L}}^{(p)} = \sum_{h =1}^{h_0} \widehat{\bm{\Sigma}}^{(p)}_{\beta}(h) \widehat{\bm{\Sigma}}^{(p)}_{\beta}(h)^\top, \quad \widehat{\bm{\Sigma}}^{(p)}_{\beta}(h) = \frac{1}{T-h}\sum_{h=1}^{T-h}\left(\widetilde{\bm{\beta}}_{p,t+h}-\overline{\widetilde{\bm{\beta}}_p}\right)\left(\widetilde{\bm{\beta}}_{p,t}-\overline{\widetilde{\bm{\beta}}_p}\right)^\top,
\end{align}
where $\overline{\widetilde{\bm{\beta}}}_p = 1/T\sum_{t=1}^T\widetilde{\bm{\beta}}_{p,t}$.
Thus we estimate the $p$\textsuperscript{th} factor by:
\begin{align}\label{eq:28}
\widehat{\bm{f}}_{p,t} = \widehat{\bm{A}}_p^\top\widetilde{\bm{\beta}}_{p,t}.
\end{align}

The estimated dimension reduced FPC scores are
\begin{align*}
\widehat{\bm{\beta}}_{p,t} = \widehat{\bm{A}}_p \widehat{\bm{f}}_{p,t}.
\end{align*}

The estimator for the original function $\mathcal{X}^{(i)}_t(u)$ is: 
\begin{align}\label{eq:26}
\widehat{\mathcal{X}}^{(i)}_{t}(u) = \sum_{p = 1}^{p_0} \left[\widehat{\bm{\beta}}_{p,t}\right]^{(i)} \widehat{\gamma}^{(i)}_p(u) = \sum_{p = 1}^{p_0} \left[\widehat{\bm{A}}_p\widehat{\bm{f}}_{p,t}\right]^{(i)} \widehat{\gamma}^{(i)}_p(u), \quad i = 1,\ldots, N, \quad t = 1,\ldots, T,
\end{align}
where again $[\cdot]^{(i)}$ denotes the $i$\textsuperscript{th} element of the vector. 

\subsection{Forecasting}\label{sec:2.4}

With two-fold dimension reduction, information of serial correlation is contained in the factors $\bm{f}_{p,t}$. To forecast $N$-dimensional functional time series, we could instead make forecast on the estimated factors. Scalar or vector time series models could be applied. We suggest univariate time series models: autoregressive moving average (ARMA) models, for instance, since the factors are mutually uncorrelated. Recall that we have retained $p_0$ FPC scores and $r$ factors in the estimation process. Consequently, we need to fit $r\times p_0$ ARMA models on the estimated factors $\left\{\widehat{\bm{f}}_{p,1},\ldots, \widehat{\bm{f}}_{p,t}\right\}$. The prediction of the functions could be calculated as:
\begin{align}\label{eq:27}
\widehat{\mathcal{X}}^{(i)}_{T+h|T}(u) = \sum_{p = 1}^{p_0} \left[\widehat{\bm{A}}_p\widehat{\bm{f}}_{p,T+h|T}\right]^{(i)}\widehat{\gamma}^{(i)}_p(u), \quad i = 1,\ldots, N,
\end{align}
where $\widehat{\mathcal{X}}^{(i)}_{t+h|t}(u)$ is the $h$-step-ahead forecast at time $t$, and $h$ denotes a forecast horizon.

Prediction intervals for functions could also be constructed. In this paper, we use a bootstrapping approach. The bootstrapped function forecast is
\begin{align*}
\widehat{\mathcal{X}}^{(i),b}_{T+h|T}(u) = \sum_{p = 1}^{p_0} \left[\widehat{\bm{A}}_p\widehat{\bm{f}}^b_{p,T+h|T}\right]^{(i)} \widehat{\gamma}^{(i)}_p(u) + \widehat{\epsilon}^{(i),b}_{T+h|T}(u),\qquad b = 1,\ldots,B,
\end{align*}
where $B$ is the number of bootstraps, and $\widehat{\bm{f}}^b_{p,T+h|T}$ is the bootstrapped prediction of the factor. A bootstrapped residual $\widehat{\epsilon}^{(i),b}_{T+h|T}(u)$ is added, which is resampled from $\{ \mathcal{X}^{(i)}_t(u) - \widehat{\mathcal{X}}^{(i)}_t(u)\}$. Lower and Upper prediction bounds are calculated as the $100\times(\alpha/2)$\textsuperscript{th} and $100\times (1-\alpha/2)$\textsuperscript{th} percentile of the bootstrapped forecasts, where $\alpha$ is the level of significance.

The bootstrapped prediction of the factors $\widehat{\bm{f}}^b_{p,T+h|T}$ can be constructed in several ways. Here in this paper, we use an intuitive approach consisting of four steps: 
\begin{enumerate}
\item
Fit an ARMA(p,q) model to the first half of the estimated factors $\widehat{f}_{p,1},\ldots, \widehat{f}_{p,T_1}$, $T_1 = T/2$, and make $h$-step ahead prediction $\tilde{f}_{p,T_1+h}$. The prediction error is $\xi_{p,1} = \widehat{f}_{p,T_1+h}-\tilde{f}_{p,T_1+h}$.
\item
Fit another ARMA(p,q) model to $\widehat{f}_{p,1},\ldots, \widehat{f}_{p,T_1+1}$, and make $h$-step ahead prediction $\tilde{f}_{p,T_1+1+h}$. The prediction error is $\xi_{p,2} = \widehat{f}_{p,T_1+1+h}-\tilde{f}_{p,T_1+1+h}$.
\item
Following the first two steps we acquire $j$ prediction errors such that $T_1+j+h= T$. Resample $\xi_p^*$ from $\{\xi_{p,1},\ldots,\xi_{p,j}\}$. 
\item
Construct the bootstrap prediction $\widehat{f}^b_{p,T+h|T} = \widehat{f}_{p,T+h|T} + \xi_p^*$, where $ \widehat{f}_{p,T+h|T}$ is the point forecast.
\end{enumerate}

The process above needs to be repeated $r$ times to generate each element in $\widehat{\bm{f}}^b_{p,T+h|T}$.

Alternatively, the bootstrapped functions can be used to construct prediction region \citep{ZP17}. Denote $q(\alpha)$ as the $\alpha$-quantile of $\big\|\widehat{\mathcal{X}}_{T+h|T}^{i,b}-\widehat{\mathcal{X}}_{T+h|T}^{i}\big\|_2$. Then the $(1-\alpha)\times100\%$ bootstrap predictive region consists of all $\mathcal{X}$ such that 
\begin{align*}
\left\|\mathcal{X}-\widehat{\mathcal{X}}_{T+h|T}^{(i)}\right\|_2 \le q(\alpha).
\end{align*}

\section{Asymptotic Theory}\label{AT}

We derive consistency results for dimension reduced estimates of the functions. Lemmas and detailed proofs are provided in the Appendix. 
First, some assumptions are made on the functional time series and the FPCA estimation process.
\begin{assumption}\label{as:1}
For each $i = 1,\ldots, N$, the functions $\{\mathcal{X}^{(i)}_t(u), t\in \mathbb{Z}\}$ are stationary and $L^4$-m-approximable, which also satisfies 
\begin{align*}
\rm \sup_t \|\mathcal{X}^{(i)}_t(u)\| < \infty
\end{align*}
\end{assumption}

\begin{assumption}\label{as:7}
For each $i = 1,\ldots, N$, the $h$ lag covariance operator satisfies
\begin{align*}
\sum_{h=-\infty}^\infty \left\|C^{(i)}_h\right\|_\mathcal{S} < \infty,
\end{align*}
where $\|\cdot\|_\mathcal{S}$ denotes the Hilbert-Schmidt norm.
\end{assumption}

In Assumption~\ref{as:1}, it is assumed the dependence structure of the functional time series for each population. The definition of $L^p$-m-approximable in Assumption~\ref{as:1} and the Hilbert-Schmidt norm in Assumption~\ref{as:7} can be found in the Appendix. The article also provides a simple sufficient condition for Assumption~\ref{as:7} to hold, as in the following proposition. 
\begin{proposition}
If $\mathcal{X}_t$ is $L^2$-m-approximate, then Assumption~\ref{as:7} holds.
\end{proposition}

An example of functional time series that satisfies this assumption is the simplest functional AR(1) model. The proof of the proposition and that functional AR(1) is $L^2$-m-approximate is included in the Appendix.

\begin{assumption}\label{as:2}
For each $i = 1, \ldots, N$, the eigenvalues $\lambda_p^{(i)}, p = 1,\ldots$ are distinct.
\end{assumption}

\begin{assumption}\label{as:3}
In the estimation of the functional principal components, for each population $i$, the empirical eigenfunctions are in the same direction of the true function, i.e.,$\langle \gamma_p^{(i)}, \widehat{\gamma}_p^{(i)}\rangle > 0$
\end{assumption}

\begin{assumption}\label{as:4}
In the estimation of the functional principal components, for each population $i$, the bandwidth parameter $q^3 = o\left(T/N\right)$, and $q \rightarrow \infty$.
\end{assumption}

 Assumption~\ref{as:2} is a very common assumption in FPCA. Further, Assumption~\ref{as:3} ensures that we choose the correct sign for each eigenfunction. This assumption is used to serve for theoretical proof. In practice, the sign of the estimated eigenfunction does not make a difference because the problem vanishes once we take the product of the estimated eigenfunction and the corresponding estimated FPC score. Assumption~\ref{as:4} imposes a condition on the rate of the bandwidth parameter $q$, which has been previously defined in~\eqref{eq:7}. The two conditions in Assumption~\ref{as:4} also imply $N = o(T)$, that is the number of populations grows not as rapidly as the sample size. 

The following assumptions are made on the second dimension reduction step, the factor model as stated in \eqref{eq:5}. As in Section~\ref{sec:2.2}, in the following, we again omit the subscript $p$ for conciseness.  First, let's define some notations. We use $\|\bm{M}\|_2$ to denote the $L_2$ norm of the matrix or vector. When $\bm{M}$ is a matrix, it is the greatest singular value. We use $\|\bm{M}\|_{\min}$ to denote the smallest singular value. We use $a \asymp b$ to denote $\{a = O(b)\}\cap \{b=O(a)\}$, that is $a$ and $b$ are of the same order.

\begin{assumption}\label{as:5}
For $p$ in 1 to $p_0$, $\|\bm{\Sigma}^{(p)}_{f}(h)\|_2 \asymp N^{1-\delta} \asymp \|\bm{\Sigma}^{(p)}_f(h)\|_{\min}$, where $0 \le \delta < 1$.

\end{assumption}

\begin{assumption}\label{as:6}
For $p$ in 1 to $p_0$, $\|\bm{\Sigma}^{(p)}_{e,f}(h)\|_2 = O(\|\bm{\Sigma}^{(p)}_f(h)\|_2)$, and $\|\bm{\Sigma}^{(p)}_{e}(h)\|_2 = O(\min\{NT^{-1/2}, 1\})$.
\end{assumption}

In Assumption~\ref{as:5}, it is assumed that the order of the lag covariance of factor $\bm{f}_{p,t}$ is related to the dimension of $\bm{\beta}_{p,t}$ by a factor $\delta \in [0,1)$. In Assumption~\ref{as:6}, it is assumed that the strength of the lag cross-covariance between factors and errors is not bigger than that of the lag covariance of the factors, and that the lag covariance of the error term is at least bounded or of constant rate. Since what the model does essentially is principal component analysis, we want to ensure that most of the covariation of $\bm{\beta}_{p,t}$ is contained in the lower dimension factors $\bm{f}_{p,t}$.

\begin{assumption}\label{as:9}
\begin{align*}
\|\theta_t^{(i)}(u)\|_2 = o_P(1), \qquad N \rightarrow \infty
\end{align*}
where $\theta_t^{(i)}(u)$ is defined in \eqref{eq:21}.
\end{assumption}

\begin{assumption}\label{as:11}
For each i, 
\begin{align*}
[\bm{e}_{p,t}]^{(i)} = O_P\left(\frac{1}{\sqrt{N}}\right), \qquad N,T \rightarrow \infty,
\end{align*} 
where $\bm{e}_{p,t}$ is defined in \eqref{eq:5}, and $[\bm{e}_{p,t}]^{(i)}$ denotes the $i$th element in the vector $\bm{e}_{p,t}$.
\end{assumption}

In Assumption~\ref{as:9} and~\ref{as:11}, it is assumed the error terms in both dimension reduction steps to be small, which is a natural assumption in principal component analysis.

\begin{theorem}\label{th:1}
Under Assumptions~\ref{as:1} to \ref{as:4}, $\left\|\widetilde{\bm{\beta}}_{p,t}-\bm{\beta}_{p,t}\right\|_2$ converges to zero in probability as $N, T \rightarrow \infty$, where the vectors $\bm{\beta}_{p,t}$ and $\widetilde{\bm{\beta}}_{p,t}$ are defined in~\eqref{eq:4} and~\eqref{eq:6}. 
\end{theorem}
\begin{theorem}\label{th:2}
Under Assumptions~\ref{as:1} to \ref{as:6}, assuming $N^\delta T^{-1/2} = o(1)$, we have
\begin{align*}
\|\widehat{\bm{A}}_p-\bm{A}_p\|_2 = O_P\left(N^{\delta} T^{-1/2}+N^{\delta-1}\right)= o_P(1)
\end{align*}
\end{theorem}

In Theorem~\ref{th:2}, we have proved the convergence rate of the estimated factor loadings. When $\delta$ is 0, the convergence rate becomes $\left(T^{-1/2} + N^{-1}\right)$, which is quite fast. However when $\delta$ is close to 1, the rate of convergence is very slow. 

\begin{theorem} \label{th:3}
Under Assumptions~\ref{as:1} to \ref{as:6},
\begin{align*}
\frac{1}{N}\left\|\widehat{\bm{\beta}}_{p,t}-\bm{\beta}_{p,t}\right\|_2 = O_P\left(N^{(\delta-1)/2}T^{-1/2}\right) + O_P\left(\frac{1}{N}\right).
\end{align*}
\end{theorem}

\begin{theorem}\label{th:4}
Under Assumptions~\ref{as:1} to \ref{as:11}, assuming $N^{\delta/2}T^{-1/2} = o(1)$,
\begin{align*}
\frac{1}{N}\sum_{i=1}^{N} \left\|\widehat{\mathcal{X}}^{(i)}_t(u)-\mathcal{X}^{(i)}_t(u)\right\| = o_P(1), \qquad N, T \rightarrow \infty
\end{align*}
\end{theorem}

Theorem~\ref{th:3} states the convergence rate for the estimated FPC scores $\bm{\beta}_{p,t}$. Theorem~\ref{th:4} proves that the approximated functions are good estimates of the true functions. The rate of convergence is calculated in the Appendix. 

We also investigates the prediction error of the model. After dimension reduction, classic time series models are fitted to the estimated factors $\widehat{\bm{f}}_{p,t}$. In this paper, we use the AR(1) model as an example in the proof for asymptotic property. Let $f^{(i)}_{p,t}$ denote the $i$th element in the vector $\bm{f}_{p,t}$. The AR(1) model is 
\begin{align*}
f^{(i)}_{p,t} = \phi_{p,i}f^{(i)}_{p,t-1} + \omega^{(i)}_{p,t}, \quad t = 2,\ldots,
\end{align*}
where $\phi_{p,i}$ is the AR coefficient which satisfy $|\phi_{p,i}| < 1$, and $\omega^{(i)}_{p,t}$ is the white noise. We define $\Gamma = \max_{p,i}(\Gamma_{p,i})$, where $\Gamma_{p,i} = \left|\sum_{j=1}^{h-1}\phi^j_{p,i}\omega^{(i)}_{p,T+h-j}\right|$. We have the following theorem.

\begin{theorem}\label{th:5}
Under Assumptions~\ref{as:1} to \ref{as:11}, assuming $N^{\delta/2}T^{-1/2} = o(1)$,
\begin{align*}
\frac{1}{N}\sum_{i=1}^N\left\|\widehat{\mathcal{X}}^{(i)}_{T+h|T}(u)-\mathcal{X}_{T+h}^{(i)}(u)\right\| = o_P(1) + O_P(N^{-1/2}\Gamma), \qquad N,T \rightarrow \infty,
\end{align*}
\end{theorem}
 We see that the forecast error includes a component that converges to zero which comes from the estimation error, and a component that is $O_P(N^{-1/2}\Gamma)$ which measures the error from the forecast model. In the ordinary setting of univariate AR(1) model, $\Gamma = O_P(1)$. So the second part also converges, that is $N^{-1/2}\Gamma = o_P(1)$.

\section{Simulation Studies}\label{SS}
We illustrate our method using simulated data. We compare results using the proposed high-dimensional functional time series (HDFTS) model and a univariate functional time series (FTS) model where each population is modeled and predicted as a independent functional time series. 
\subsection{Data generation}
We generate $N$ populations of functional time series data.
The $i$\textsuperscript{th} function at time $t$ is constructed by
\begin{align*}
\mathcal{X}^{(i)}_t(u) = \sum_{p=1}^{2}\beta_{p,t}^{(i)} \gamma_p^{(i)}(u) + \theta_t^{(i)}(u), \qquad t = 1,\ldots, T, \quad i = 1,\ldots, N
\end{align*}
where $\theta_t^{(i)}(u) = \sum_{p=3}^\infty \beta_{p,t}^{(i)} \gamma_p^{(i)}(u)$.

The coefficients $\beta_{p,t}^{(i)}$ for all $N$ populations are combined and generated by
\begin{align*}
\bm{\beta}_{p,t} = \bm{A}_{p}  \bm{f}_{p,t}, \qquad p = 1,\ldots,\infty,
\end{align*}
where $\bm{\beta}_{p,t} = \{\beta_{p,t}^1,\ldots, \beta_{p,t}^N\}$. $\bm{A}_p$ is a $N \times N$ matrix, and $\bm{f}_{p,t}$ is a $N \times 1$ vector.

We assume that $\beta^{(i)}_{p,t}$ have mean 0 and variance 0 when $p > 3$, so we only construct the coefficients $\bm{\beta}_{p,t}$ for $p=1, 2, 3$.

The first set of coefficients $\bm{\beta}_{1,t}$ for $N$ populations are generated with $\bm{\beta}_{1,t} = \bm{A}_{1}  \bm{f}_{1,t}$. Each element in matrix $\bm{A}_1$ is generated by $a_{ij} = N^{-1/4}\times b_{ij}$, where $b_{ij}\sim \mathcal{N}(2,4)$.

The factors $\bm{f}_{1,t}$ are generated using autoregressive model of order 1 (AR(1)). Define the $i$\textsuperscript{th} element in vector $\bm{f}_{1,t}$ as $f_{1,t}^{(i)}$. Then, $f_{1,t}^1$ is generated by $f_{1,t}^1 = 0.5 f_{1,t-1}^1+ \omega_t$, where $\omega_t$ are independent $\mathcal{N}(0,1)$ random variables. We generate $f_{1,t}^{(i)}, \ i = 2,\ldots N$ by $f_{1,t}^{(i)} = (1/N) g_t^{(i)}$, where $g_t^{(i)}, i=2,\ldots, N$ are also AR(1) and follow $g_t^{(i)} = 0.2g_{t-1}^{(i)} + \omega_t$. It is ensured that most of the variance of $\bm{\beta}_{1,t}$ can be explained by one factor.
The second coefficient $\bm{\beta}_{2,t}$ are constructed the same way as $\bm{\beta}_{1,t}$.

We also generate the third FPC scores $\bm{\beta}_{3,t}$ but with small values. $\bm{A}_3$ is generated by $a_{ij} = N^{-1/4}\times b_{ij}$, where $b_{ij}\sim \mathcal{N}(0,0.04)$. The factors $\bm{f}_{3,t}$ are generated as $\bm{f}_{1,t}$.

The three basis functions are constructed by $\gamma_1^{(i)}(u) = \sin(2\pi u+ \pi i/2)$,  $\gamma_2^{(i)}(u) = \cos(2\pi u + \pi i/2)$ and $\gamma_3^{(i)}(u) = \sin(4\pi u+ \pi i/2)$, where $u \in [0,1]$. The functional time series for the $i$\textsuperscript{th} population is constructed by
\begin{align*}
\mathcal{X}_t^{(i)}(u) = \left[\bm{\beta}_{1,t}\right]_i \gamma_1^{(i)}(u) + [\bm{\beta}_{2,t}]_i \gamma_2^{(i)} +  [\bm{\beta}_{3,t}]_i \gamma_3^{(i)},
\end{align*}
where $[\cdot]_i$ denotes the $i$\textsuperscript{th} element of the vector.

\subsection{Model fitting}
Simulated data are generated under different settings of $N$ and $T$ values. The proposed model is fitted to the data. The bandwidth parameter is simply chosen as $\sqrt{T}$ in each case. We use fraction of variation explained (FVE) to choose both the number of FPC score $\widehat{p}_0$ and the number of retained factors $
\widehat{r}$. We require the first $\widehat{p}_0$ FPC scores to explain 99\% of each population of functional time series, and the first $\widehat{r}$ factors to explain also 99\% of each FPC scores. For different populations, the chosen number of FPC scores can be different according to FVE. Our model requires to select the same number of FPC scores for each population. Therefore, we choose $\widehat{p}_0$ to be the largest number of FPC scores needed. The number of retained factors $\widehat{r}$ in the factor models can also be different for each $p = 1, \ldots, \widehat{p}_0$. Therefore, we use $\widehat{r}_1,\ldots,\widehat{r}_{p_0}$ to denote the number of factors chosen for each FPC score. 
 
We first look at the estimation error under different settings. The estimation error is calculated using mean norm of residuals (MNR):
\begin{align*}
MNR = \frac{1}{NT}\sum_{i=1}^N\sum_{t=1}^T \sqrt{\sum_{j=1}^w [\mathcal{X}_t^{(i)}(u_j)-\widehat{\mathcal{X}}_t^{(i)}(u_j)]^2},
\end{align*}
where $\mathcal{X}_t^{(i)}(u_j)$ denotes the function value at discrete time point $u_j,$ for $j = 1\ldots,w$, and $w$ is the total number of discrete points in $[0,1]$. We use $w=51$ throughout simulation study.

Next we compare the forecast performance of our model with the independent forecast model. The independent forecast model follows similar idea of the approach in \cite{HU07} for each population. When fitting models to the data, we use expanding window prediction. The simulated data is divided into a training set with size $T_1$ and a test set with size $T_2$, and $T_1+T_2 = T$. In this study, we use $T_2 = (1/4)\times T$. The proposed models are fitted to the training set and forecasts are made based on fitted models. Then the test set is used for forecast evaluation. Each time we increase the training size by one and refit the model. New forecasts are made each time. Finally all the prediction errors for one-, two-, and three-step-ahead forecasts are collected and means are taken.

We use discretized mean absolute forecast error (MAFE) and mean squared forecast error (MSFE). 
\begin{align*}
\begin{split}
\mbox{MAFE}(h) &= \frac{1}{(T_2+1-h)\times w}\sum_{\eta=h}^{T_2} \sum_{j=1}^{w} \Big | \mathcal{X}_{T_1+\eta}(u_j)-\widehat{\mathcal{X}}_{T_1+\eta|T_1+\eta-h}(u_j)\Big |,\\
\mbox{MSFE}(h) &= \frac{1}{(T_2+1-h)\times w}\sum_{\eta=h}^{T_2}\sum_{j=1}^{w} \left[\mathcal{X}_{T_1+\eta}(u_j)-\widehat{\mathcal{X}}_{T_1+\eta|T_1+\eta-h}(u_j)\right]^2,
\end{split}
\end{align*}
where $\widehat{\mathcal{X}}_{T_1+\eta|T_1+\eta-h}$ represents the $h$-step-ahead prediction using data $t = 1,\ldots, T_1+\eta-h$ fitted in the model, and $\mathcal{X}_{T_1+\eta}(u_j)$ denotes the holdout function.

For each combination of $N$ and $T$ values, we replicate the simulation 100 times, and calculate the mean of the errors MNR, MAFE and MSFE.

\subsection{Results}
The estimation errors of the proposed model are presented in Table~\ref{tab:3}. With the increase of $N$ and $T$, the MNR becomes smaller, which can be seen as a concordant with Theorem \ref{th:4}. Under all four settings, we select the number of FPC scores to be two. The number of factors selected for the first and second FPC scores $\widehat{r}_1$ and $\widehat{r}_2$ are different in each simulation, but mostly equal to two or three.
\begin{table}[!htbp] 
\tabcolsep 0.6in
\centering 
\caption{The MNR under different settings} \label{tab:3} 
\begin{tabular}{@{}lcc@{}} 
\toprule
$(N, T)$   &  MNR & $\widehat{p}_0$ \\
\midrule
$(20, 20)$ & $1.756$ & $2$ \\ 
($40, 50$)  & $1.226$ & $2$ \\ 
($60, 80$) & $0.804$ & $2$ \\ 
($100, 150$)  & $0.622$ & $2$ \\ 
\bottomrule
\end{tabular} 
\end{table} 
			
Table~\ref{tab:2} shows the sample mean of the MAFE and MSFE in different settings. Each number in the table is the mean of the $h$-step-ahead errors taken over all $N$ populations. The smaller value is in bold face. It can be seen that the proposed model produces smaller forecast errors in almost all settings and all forecast horizons.

\begin{table}[!htbp] 
\tabcolsep 0.34in
\centering 
\caption{The mean MAFE and MSFE values when fitting the independent functional time series model and the proposed high-dimensional functional time series model two models for one-, two-, and three-step-ahead forecasts} \label{tab:2} 
\begin{tabular}{@{}lcccccc@{}} 
\toprule
 & & \multicolumn{2}{c}{MAFE}  && \multicolumn{2}{c}{MSFE}\\
$(N, T)$ & $h$ & FTS & HDFTS && FTS & HDFTS\\
\midrule
& $1$ & $1.136$ & $\bm{1.134}$ && $\bm{2.595}$ &$2.597$ \\ 
$(20, 20)$ & $2$ & $1.310$ & $\bm{1.266}$& &  $3.539$ &$\bm{3.256}$  \\ 
& $3$ & $1.380$ & $\bm{1.324}$ & &$3.983$ &$\bm{3.630}$  \\ 
\midrule
& $1$ & $0.917$ & $\bm{0.872}$ & &$1.688$ &$\bm{1.532}$  \\ 
($40, 50$) & $2$ & $1.038$ & $\bm{0.970}$& &$2.187$ &$\bm{1.908}$  \\ 
& $3$ & $1.096$ & $\bm{0.998}$ & &$2.438$ &$\bm{2.005}$  \\ 
\midrule
& $1$ & $0.817$ & $\bm{0.763}$ && $1.327$ &$\bm{1.163}$  \\ 
($60, 80$) & $2$ & $0.935$ & $\bm{0.858}$ &&$1.752$ &$\bm{1.473}$  \\ 
& $3$ & $0.980$ & $\bm{0.877}$ && $1.942$ &$\bm{1.540}$  \\ 
\midrule
& $1$ & $0.688$ & $\bm{0.644}$ && $0.992$ &$\bm{0.863}$  \\ 
($100, 150$) & $2$ & $0.799$ & $\bm{0.737}$ && $1.331$ &$\bm{1.128}$  \\ 
& $3$ & $0.856$ & $\bm{0.764}$ & &$1.520$ &$\bm{1.231}$  \\ 
\bottomrule
\end{tabular} 
\end{table} 
		
\section{Empirical Studies}\label{MF}

We also illustrate our method using an empirical data set. The Japanese sub-national mortality rates in 47 prefectures are used to demonstrate the effectiveness of our proposed method. Available from the \cite{JMD}, the data set contains yearly age-specific mortality rates over a span of 41 years from 1975 to 2015. The observations are the yearly mortality curves from ages 0 to 110 years, where age is treated as the continuum in the rate function. In this study, the data at ages 95 and older are grouped together, to avoid problems associated with erratic rates at these ages.

\begin{figure}[!htbp]
\centering
\includegraphics[width=12cm]{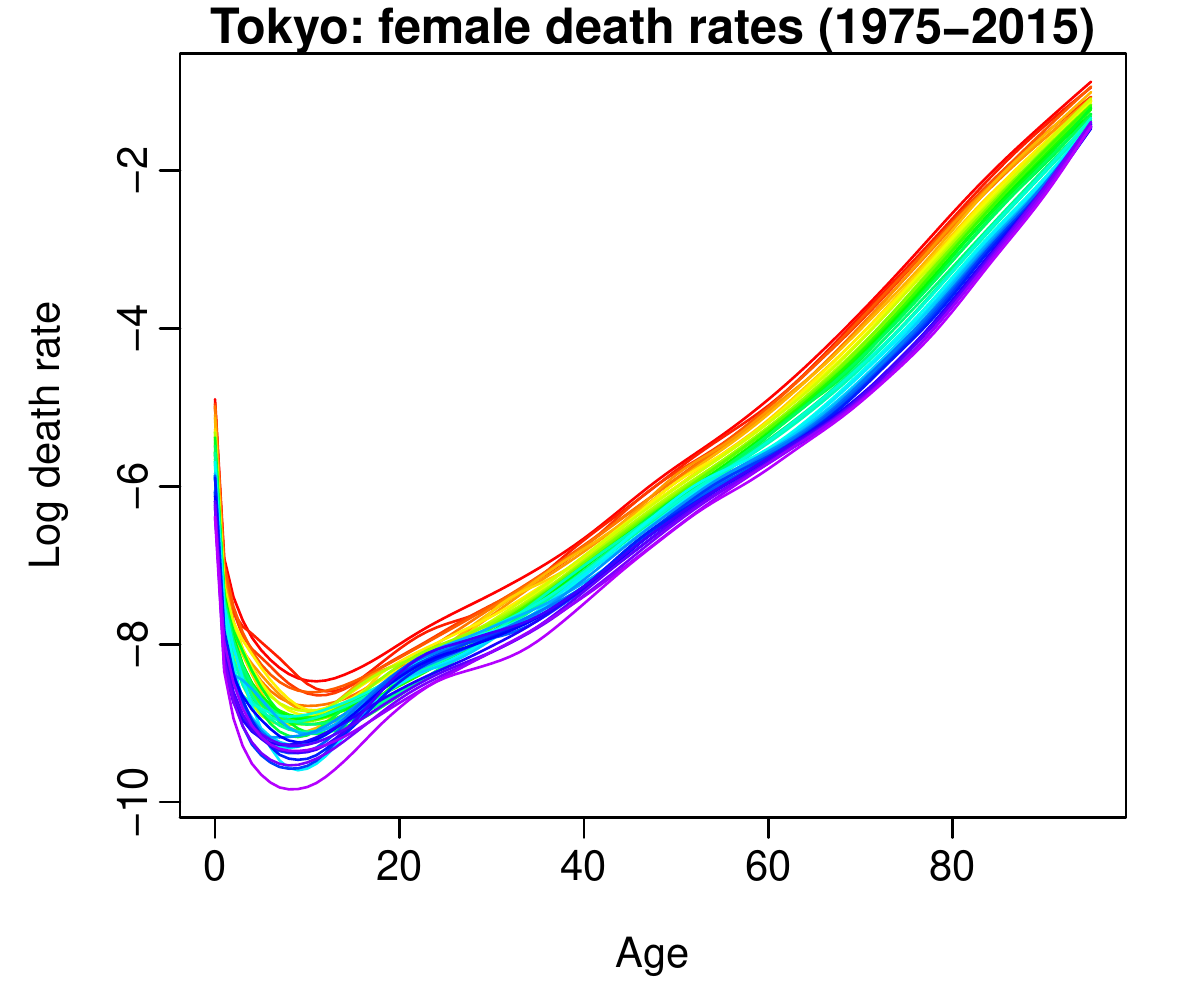} 
\caption{Log smoothed female mortality in the Tokyo prefecture from 1975 to 2015}\label{fig:1}
\end{figure}

A graphical display of the functional time series is presented in Figure~\ref{fig:1}. The figure presents the log smoothed female age-specific mortality rates in the Tokyo prefecture, where the red lines represent more distant data and the purple lines represent more recent years. The curves are smoothed using penalized regression splines with a monotonically increasing constraint after the age of 65 \citep[see][]{Wood94, HU07}. 

 The dimension of the functional time series is $N = 47$, which is greater than the sample size $T= 41$. With the two-fold dimension reduction model, we use the first three FPC scores for each population, and the first three factors. 

The expanding window approach is again used as described in the simulation study. The first 26 years of data (from 1975 to 2000) are allocated to the training set, and the last 15 years of data (from 2001 to 2015) are allocated to the testing set. In each fitting process, the bandwidth parameter $q$ is chosen as $\sqrt{T^*}$, where $T^*$ is the number of years fitted to the model. We compare the forecast accuracy of our proposed method with the independent functional time series model, where each sub-national population is forecast individually using the approach in \cite{HU07}. With Hyndman and Ullah's mortality model, the FPCA is performed on the functional time series of each prefecture and FPC scores are estimated and fitted to classical time series model to make predictions. Then functional forecast are produced using predicted scores. In this method, the prediction intervals are constructed by calculating the total variance of the principal components. This independent forecast model does not take into account the dependence between the sub-national populations. 

For female mortality rates, prediction errors are calculated and show that the proposed method outperforms the independent model in general. Specifically, in 24 out of 47 prefectures, the proposed model produces smaller MAFE in all one to ten forecast horizons. In 45 out of 47 prefectures, the proposed model produces smaller mean MAFE taken across all forecast horizons. 

We have also fitted the male mortality data of the 47 prefectures. The female and male forecast errors are summarized in Table~\ref{tab:1}. Each number in the table is the mean error taken across 47 prefectures.  FTS stands for the alternative independent Hyndman and Ullah method and HDFTS stands for the proposed high-dimensional functional time series model. For female data, our model outperforms independent FTS model in both MAFE and MSFE values. For male data, however, our model produces smaller MAFE but does not have an advantage in MSFE values.

\begin{table}[!htbp] \centering 
\tabcolsep 0.135in
  \caption{MAFE and MSFE for the Japanese female and male rates} \label{tab:1}
\begin{tabular}{@{}lccccccccc@{}} 
\toprule
&\multicolumn{4}{c}{Female} & & \multicolumn{4}{c}{Male}\\
& \multicolumn{2}{c}{MAFE} & \multicolumn{2}{c}{MSFE} & &\multicolumn{2}{c}{MAFE} & \multicolumn{2}{c}{MSFE} \\
$h$ & FTS & HDFTS  & FTS  & HDFTS && FTS &HDFTS & FTS & HDFTS \\ 
\midrule
$1$ & $0.174$ & $\bm{0.164}$ &$0.293$&$\bm{0.286}$&& $0.266$ & $\bm{0.261}$ &$\bm{0.609}$&$0.619$\\ 
$2$ & $0.179$ & $\bm{0.165}$ &$0.315$&$\bm{0.293}$&& $0.274$ & $\bm{0.261}$ &$0.634$&$\bm{0.607}$\\ 
$3$ & $0.182$ & $\bm{0.168}$ &$0.323$&$\bm{0.285}$ && $0.280$ & $\bm{0.274}$ &$\bm{0.646}$&$0.673$\\ 
$4$ & $0.186$ & $\bm{0.170}$ &$0.336$&$\bm{0.302}$ && $0.291$ & $\bm{0.281}$ &$\bm{0.673}$&$0.684$\\ 
$5$ & $0.187$ & $\bm{0.169}$ &$0.334$&$\bm{0.310}$ && $0.294$ & $\bm{0.280}$ &$\bm{0.662}$&$0.691$\\ 
$6$ & $0.197$ & $\bm{0.175}$ &$0.373$&$\bm{0.337}$ && $0.311$ & $\bm{0.293}$ &$\bm{0.714}$&$0.758$\\ 
$7$ & $0.207$ & $\bm{0.174}$ &$0.406$&$\bm{0.343}$&& $0.325$ & $\bm{0.300}$&$\bm{0.749}$&$0.801$ \\ 
$8$ & $0.217$ & $\bm{0.178}$ &$0.441$&$\bm{0.365}$ && $0.342$ & $\bm{0.314}$ &$\bm{0.809}$&$0.860$\\ 
$9$ & $0.229$ & $\bm{0.181}$ &$0.478$&$\bm{0.384}$ && $0.357$ & $\bm{0.322}$ &$\bm{0.841}$&$0.913$\\ 
$10$ & $0.232$ & $\bm{0.188}$ &$0.479$&$\bm{0.419}$&&  $0.365$ & $\bm{0.323}$&$\bm{0.838$}&$0.906$ \\ 
\midrule
Mean & $0.199$ & $\bm{0.173}$  &$0.378$&$\bm{0.332}$&& $0.311$ & $\bm{0.291}$ &$\bm{0.717}$&$0.751$\\ 
Median & $0.192$ & $\bm{0.172}$ &$0.354$&$\bm{0.323}$&&  $0.302$ &  $\bm{0.287}$&$\bm{0.693}$&$0.724$ \\
\bottomrule

\end{tabular} 
\end{table} 

\begin{figure}[!htbp]
\centering
\includegraphics[width=12cm]{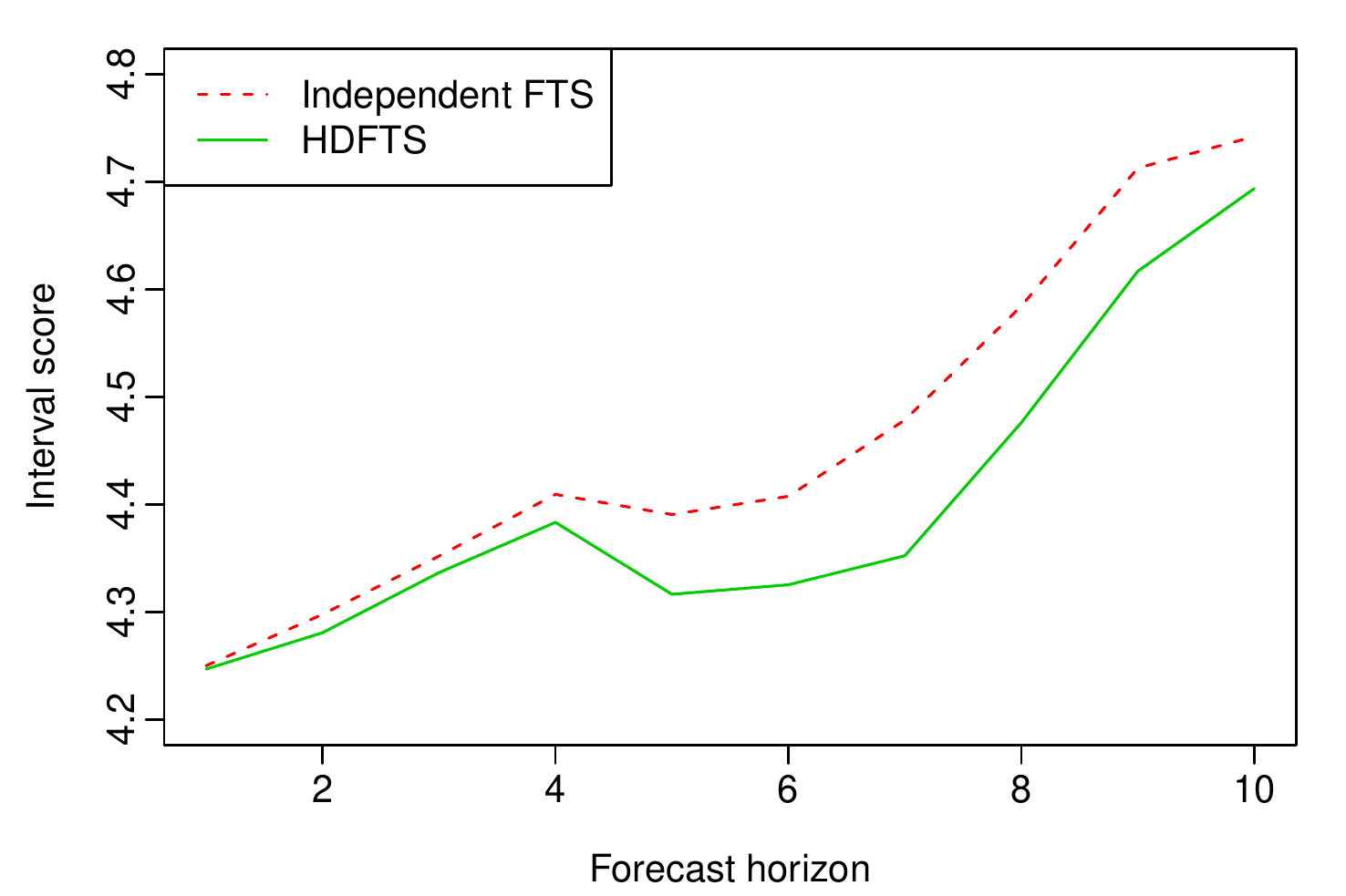} 
\caption{Mean interval score for one- to ten-step-ahead forecast. The green solid line represents the mean interval score for the high-dimensional functional time series model. The red colored dotted line represents the mean interval score for independent functional time series forecast}\label{fig:2}
\end{figure}

Prediction intervals are calculated based on the bootstrap approach. We use interval score as an evaluation for interval forecast. Let $\widehat{\mathcal{X}}^u_{n+h|n}$ and $\widehat{\mathcal{X}}^l_{n+h|n}$ denote the upper and lower $(1-\alpha)\times100\%$ prediction bounds, and $\mathcal{X}_{n+h}$ is the realized value. The discretized interval score at point $u_j$ is defined as
\begin{align*}
S_{\alpha}(u_j) &= \left[\widehat{\mathcal{X}}^u_{n+h|n}(u_j)-\widehat{\mathcal{X}}^l_{n+h|n}(u_j)\right] \nonumber\\
&+\frac{2}{\alpha}\left[\widehat{\mathcal{X}}^l_{n+h|n}(u_j)-\mathcal{X}_{n+h}(u_j)\right]\mathbb{1}\left\{\mathcal{X}_{n+h}(u_j)<\widehat{\mathcal{X}}^l_{n+h|n}(u_j)\right\}\\
&+\frac{2}{\alpha}\left[\mathcal{X}_{n+h}(u_j)-\widehat{\mathcal{X}}^u_{n+h|n}(u_j)\right]\mathbb{1}\left\{\mathcal{X}_{n+h}(x_j)>\widehat{\mathcal{X}}^u_{n+h|n}(u_j)\right\},
\end{align*}
where $\alpha$ is the level of significance, and $\mathbb{1}\{\cdot\}$ is a binary indicator function. According to this standard, the best predicted interval is the one that gives the smallest interval score. In the functional case here, the point-wise interval scores are computed and the mean over the discretized ages is taken as a score for the whole curve. Then the average scores over all populations are calculated. Mean interval scores are shown in Figure~\ref{fig:2}. Though the values are not different by large scales, the proposed high-dimensional FTS model has an apparent advantage in interval predictions especially in long-run forecast.

\section{Conclusions}\label{C}

We have proposed a two-fold dimension reduction model for modeling and forecasting high-dimensional functional time series. Our approach utilizes dynamic FPCA and factor model to represent original data with low-dimensional time series. This offers a solution to the issue of the curse of dimensionality in high-dimensional data settings. We have also provided the asymptotic properties of the model when the dimension of the functional data $N$ grows with the sample size $T$. When the tail terms in both dimension reduction steps converge to zero, the estimation error can be proved to converge to zero. Compared to the existing forecasting approaches, the proposed method has been proven to perform well both in simulations and in an empirical data analysis.

\section*{Acknowledgments}

We acknowledge the insightful comments and suggestions from the session participants of advances in mortality modeling at the International Population Conference in November, 2017. The first author would like to acknowledge financial support from a postgraduate scholarship at Australian National University.

\newpage
\bibliographystyle{myjmva}
\bibliography{fvecm}

\end{document}